\documentstyle[mprocl]{article}

\def\Journal#1#2#3#4{{#1} {\bf #2}, #3 (#4)}

\def\NPB{{\em Nucl. Phys.} B}
\def\PLB{{\em Phys. Lett.}  B}

\def\PRD{{\em Phys. Rev.} D}

\def\be{\begin{equation}}
\def\ee{\end{equation}}
\def\bea{\begin{eqnarray}}
\def\eea{\end{eqnarray}}

\begin{document}

\title{LOOP CORRECTIONS FOR 2D HAWKING RADIATION}

\author{A. MIKOVI\'C }

\address{Departamento de Fisica Teorica, Universidad de Valencia\\
46100 Burjasot, Valencia, Spain\\and\\Institute of Physics, P.O.Box 57, 
11001 Belgrade, Yugoslavia }

\author{V. RADOVANOVI\'C }

\address{Faculty of Physics, P.O.Box 368, 11001 Belgrade, Yugoslavia}


\maketitle\abstracts
{We describe how the reduced phase space quantization of the 
CGHS model of 2d black hole formation allows one to calculate the  
backreaction
up to any finite order in matter loops. We then analyze the backreaction in
the Hawking radiation up to two loops. At one-loop order the Hawking 
temperature does not change, while at two-loop order the temperature
increases three times with
respect to the classical value. We argue that such a  behavior is consistent
with the behaviour of the operator quantization Hawking flux for times which 
are not too late.
For very late times the two-loop flux goes to zero which indicates that the
backreaction can stop the black hole evaporation.}
  

Two-dimensional dilaton gravity models have turned out to be excellent toy
models of black hole formation and evaporation \cite{rev}. The generic model
is given by
\begin{equation}
S = \int_M d^{2} x \sqrt{-g}e^{-2\phi} [R + \alpha (\nabla\phi)^{2}
+ U(\phi)]  -\frac12\int_M d^{2} x \sqrt{-g}e^{-2\beta\phi}(\nabla f)^{2}
\quad,\label{1}
\end{equation}
where the 2d manifold $M$ is topologically ${\bf R}^{2}$, $g_{\mu\nu}$ is a 
Lorenzian
metric, $R$ and $\nabla$ are the corresponding curvature and covariant 
derivative, $\phi$ is the dilaton scalar field, $f$ is the matter scalar
field and $\alpha$ and $\beta$ are constants. 
The class of models described by (1) is relevant for 4d black holes
since a spherical matter collapse can be described by a model (1)
where $\alpha = 2$, $U(\phi)= 2 \exp (2\phi)$ and $\beta = 0$ for null-dust
while $\beta = 1$ for minimally coupled scalar field. As far as the 2d black
holes are concerned, the most exetnsively studied model is the CGHS model
\cite{cghs}, where $\alpha = 4$, $U(\phi)= 4\lambda^{2} = const.$ and
$\beta = 0$. This model is classically solvable \cite{cghs}, which greatly 
facilitates the computation of quantum properties \cite{m95,m96,mr}. 

The general solution for the CGHS model can be expressed in the Kruskal gauge
$\rho = \phi$, where $\rho$ is the conformal factor 
($ds^{2} = -e^{2\rho} dx^{+}dx^{-}$), as
\be
\begin{array}{rcl}
e^{-2\rho} &=& e^{-2\phi} = M_0 - \lambda^{2} x^{+}x^{-} - F_+ - F_- \\
F_p & =& \int_{-\infty}^{x^{p}} dy (x^{p} - y) T_{pp} (y)\quad,\quad 
p\in\{ +,-\} \\
T_{pp} &=& \frac12  \left( {\partial f\over\partial x^{p}}\right)^{2} \\
f &=& f_+ + f_- \quad,\quad 
f_p =\frac1{\sqrt{2\pi}} \int_0^{\infty}\frac{dk}{\sqrt k} 
( a_{pk}e^{-ikx^{p}} + c.c. )\quad.
\end{array}\label{2}
\ee
When $f_- =0$, the
solution (2) describes the collapse of left-moving matter which forms a 2d
black hole \cite{cghs}. The ADM mass of this black hole is given by
$ M = M_0 + \lambda\int_{-\infty}^{\infty} dx^{+} x^{+} T_{++} $,
while the position of the horizon is given by
$ \Delta =  \lambda^{-2} \int_{-\infty}^{\infty} dx^{+}  T_{++} $.

It is clear from the form of the solution (2) that the independent 
dynamical degrees of freedom are those of a free scalar field $f$. 
Therfore one can quantize this model by promoting the scalar field into a
hermitian operator acting on the corresponding Fock space. Consequently
the expressions in (2) can be taken as operator valued, with an appropriate
normal ordering in $T_{pp}$. From the point of view of the canonical
quantization,
the operator expressions in (2) can be understood as the Heisenberg picture 
operators in the reduced phase space quantization \cite{m95}. The physical
Hilbert space can be written as
$ {\cal H}^{*} = {\cal H}_{M_0} \otimes {\cal H}_f $,
where ${\cal H}_{M_0} $ is the Hilbert space associated with a global degree
of freedom $M_0$ (mass of the $f=0$ solution), and ${\cal H}_{f} $ is the Fock
space of the scalar field. Since the Hamiltonian operator 
$ H = M_0 + \frac12 \int dk |k| a_k^{\dagger}a_k + c_0 $
is a Hermitian operator on ${\cal H}^{*}$, the
quantum evolution will be unitary ($c_0$ is a normal-ordering constant). 

That this unitary quantum theory can 
describe evaporating black holes can be seen by considering the time
evolution of coherent
states 
$ |\psi_0\rangle = \exp(A(f_0))|0\rangle $, where $f_0$
is the initial classical matter distribution and $|0\rangle$ is
the linear dilaton vacuum \cite{m96}. The quantum effective metric is then 
defined as
$ ds^{2}= - \langle e^{2\phi} \rangle dx^{+} dx^{-}$,
which can be evaluated perturbatively as
\be \langle e^{2\phi} \rangle = e^{2\phi_0}\sum_{n=0}^{\infty} 
e^{2n\phi_0}\langle
(\delta F)^{n}\rangle \quad,\label{9}\ee
where $\delta F = F_+ + F_- - F_0$, $e^{-2\phi_0} = 
-\lambda^{2} x^{+}x^{-} - F_0$ and
$F_0$ is an arbitrary c-number function. The expansion (3) gives quantum
perturbative corrections for a classical configuration $e^{2\phi_0}$,
where $n$ counts the matter loops. The expansion (\ref{9}) is not
reliable in the strong-coupling region where $e^{-2\phi_0}\rightarrow 0$.
Also, the effective metric makes sense only in regions where the fluctuations
in the metric operator are small.

By taking $F_0 = \langle F\rangle$, the $n =0$ term in (\ref{9}) gives a 
one-loop effective conformal factor \cite{m96}
\be e^{-2\rho}=e^{-2\phi}=
 C - \lambda^{2} x^{+}x^{-} - (k/4)\log|-\lambda^{2} x^{+}x^{-}| 
- \frac12 \int_{-\infty}^{x^{+}} dy (x^{+} -y)T_{++}^{0} (y)\quad,
\label{10}\ee 
where $k$ is a numerical constant. The solution (4)
coincides with the BPP solution obtained by solving a one-loop
effective action equations of motion \cite{bpp}. The corresponding 
geometry is well-defined in the weak-coupling region $e^{-2\phi_0} > 0$, and
it describes an evaporating black hole whose apparent horizon shrinks untill
it intersects the curvature singularity at $(x_i^{+} , x_i^{-})$. 
The solution (\ref{10})
can be continously extended in the region beyond the end-point 
($-\Delta > x^{-} > x_i^{-}$), 
where it takes a  static form ($T_{++} =0$). 
Energy conservation then requires that 
$C=k/4 [\log (k/4) - 1]$, and a negative-energy shock-wave 
(thunderpop) is emitted. There is no Hawking radiation in this region, which
indicates some sort of static remnant end-state. In the evaporating phase, the
Hawking flux is the same as for the classical solution, so that the Hawking
temperature should be $\lambda/2\pi$. This can be checked by evaluating the
Bogoliubov coefficients for the background geometry (\ref{10}) \cite{mr2}.
However, a complication arises, due to the fact that the position of the 
horizon is not well-defined, 
since it is located in the strong-coupling region.
The geometry of (\ref{10}) suggests that the classical horizon undergoes a
small shift $\Delta - |x_i^{-}|$
of the order $\Delta \exp(-M/\lambda)$. In that case one can
show that the late-time Bogoliubov coefficients give a non-Planckian spectrum 
and the corresponding Hawking flux diverges. On the other hand, if one
assumes the opposite or no horizon shift, the Bogoliubov coefficients give the
standard Planckian spectrum, which agrees with the operator quantization
Hawking flux ${\cal T}=\langle T_{\sigma\sigma}\rangle$, where $\sigma$ are 
the future null-infinity
asymptotically flat coordinates for the solution (\ref{10}).

In the two-loop case, the effective conformal factor 
for a narrow matter pulse centered around $x_0^{+}$ is given by \cite{mr}
\be e^{2\rho} = e^{2\phi} = e^{2\phi_0}\left[ 1 + e^{4\phi_0}\left( C_- +
 C_+ (x^{+} - x_0^{+})^{2}\theta (x^{+} - x_0^{+} )\right) \right]
\quad, \label{11}\ee
where $e^{-2\phi_0}$ is the one-loop solution (\ref{10}), while $C_+$ and $C_-$
are constants of motion. The corresponding geometry is well-defined in the
weak-coupling region $e^{-2\phi_0} > 0$, and there are no naked singularities
for $C_+ > 0$, which happens for the matter pulses wider than the 2d
Planck length $l_P = \lambda x_0^{+}/K$, where $K$ is the momentum cut-off
used to regularize $C_+$. If one in addition takes $C_- > 0$, which is a
matter-independent constant, the corresponding geometry does not have a
curvature singularity. 

These nice features of the two-loop geometry are reflected in
the behaviour of the 
corresponding Hawking flux $\cal T$. When $C_+ > 0$, $\cal T$ is everywhere
finite and continuous,
and for not too-late times it is close to the classical Hawking flux. For later
times, $\cal T$ increases to a maximum positive value
and then drops to a minimum negative value, from which it
goes to zero for $x^{-} = -\Delta$. The total emitted energy is positive and
finite, of order $\Delta^{2}/\lambda C_+$. The behaviour of $\cal T$ suggests
that for very late times the black hole evaporation is non-thermal and goes to
zero. The results of the Bogoliubov coefficients analysis \cite{mr2} 
are qualitative 
the same as in the one-loop case, the only difference is that the 
late-time Hawking temperature increases to $3\lambda/2\pi$, which seems to be
consistent with the late-time increase of the flux $\cal T$. For 
very late times, the Bogolibov coefficents results are not reliable, since
the background geometry is not well-defined in this region. 
However, the behaviour
of the operator quantization flux is in accord with the behaviour expected 
from general
black hole considerations. The two-loop results also indicate that higher-loop 
backreaction
corrections can remove the singularities associated with the lower-order
approximations.



\end{document}